\documentstyle[11pt,amsfonts]{article}

\newcommand{\be}{\begin{equation}}
\newcommand{\ee}{\end{equation}}
\newcommand{\bea}{\begin{array}}
\newcommand{\ea}{\end{array}}
\newcommand{\beqa}{\begin{eqnarray}}
\newcommand{\eeqa}{\end{eqnarray}}
\newcommand{\bean}{\begin{eqnarray*}}
\newcommand{\eean}{\end{eqnarray*}}

\def\up#1{\leavevmode \raise.16ex\hbox{#1}}

\newcommand{\gapproxeq}{\lower
 .7ex\hbox{$\;\stackrel{\textstyle >}{\sim}\;$}}
\newcommand{\lapproxeq}{\lower .7ex\hbox{$\;\stackrel
{\textstyle <}{\sim}\;$}}

\newcounter{appendice}

\def\thebibliography#1{{\bf REFERENCES\markboth
 {REFERENCES}{REFERENCES}}\list
 {[\arabic{enumi}]}{\settowidth\labelwidth{[#1]}\leftmargin\labelwidth
 \advance\leftmargin\labelsep
 \usecounter{enumi}}
 \def\newblock{\hskip .11em plus .33em minus -.07em}
 \sloppy
 \sfcode`\.=1000\relax}

\def\BI{{\rm 1\!l}}
\begin{document}

\begin{flushright}
DSF-T-3/05\\
SU-4252-805\\
\end{flushright} 

\centerline{ \LARGE  A Perturbative Approach to Fuzzifying  Field Theories}

\vskip 2cm

\centerline{ {\sc    A. Pinzul$^{a}$ and A. Stern$^{b,c}$ }  }

\vskip 1cm
\begin{center}
{\it a)  Department of Physics, Syracuse University,\\ Syracuse, New
York 13244-1130,  USA\\}
 {\it b) Department of Physics, University of Alabama,\\ Tuscaloosa,
Alabama 35487, USA\\} {\it c) Dip. Scienze Fisiche and INFN Sez. di
Napoli,\\ Universit\`{a} Federico II, Compl. Univ. Monte S. Angelo,
Napoli, 80126, Italy}

\end{center}

\vskip 2cm

\vspace*{5mm}

\normalsize
\centerline{\bf ABSTRACT}

We propose a procedure for computing noncommutative  corrections to
the metric tensor, and apply it to scalar field
theory written on coordinate patches of smooth manifolds.  The procedure involves finding
maps to the noncommutative plane where differentiation and integration
are easily defined, and introducing a star product.  There are star
product independent, as well as dependent, corrections.  Applying the
procedure for two different star products, we find the lowest order
fuzzy corrections to scalar field theory on a sphere which is
sterographically projected to the plane.

\vskip 2cm
\vspace*{5mm}

\newpage
\scrollmode

\section{Introduction}

Currently, field theories have been successfully written down on only
a handful of noncommuting manifolds. Besides being of intrinsic
interest, the search for new noncommutative geometries is relevant for
quantum gravity and also string theory, where one expects to have to
sum over all such geometries.  Moreover, given a field theory written
on an arbitrary curved manifold there is no canonical procedure for
making the theory noncommutative.  The inverse problem can also be
problematic.  Here one should distinguish between classical and
quantum noncommutative field theories, because the
commutative limit and the classical limit may not commute. Starting
from noncommutative quantum theory it has been noted that a phase
transition can occur in passing to the commutative theory.\cite{Gubser:2000cd}
For such theories the commutative limit is  singular. On the other hand,  it should
always be possible to take the commutative limit of classical noncommutative field theory by simply setting the
noncommutative parameter to zero.

 In this article we shall be concerned with  classical
noncommutative field  theory, with  the aim of developing
 a systematic  procedure for computing noncommutative or fuzzy corrections to
classical field theories written on coordinate patches of arbitrary smooth manifolds.  We
report on progress in this direction for the case of scalar field
theory in two dimensions.
The approach taken  does not involve the more formal aspects of
noncommutative geometry.  We start   with some
noncommutative associative algebra ${\cal A}$.  There are a number of
obstacles  to constructing a noncommutative space from ${\cal A}$. In
addition to insuring that the Jacobi identity is satisfied, there is
the problem of defining the notion of derivations and integration on
this space. As we shall only be interested in two-dimensional theories
the Jacobi identity is trivially satisfied. Two-dimensional examples
where derivations and integration can be defined are the
noncommutative plane, noncommutative torus, fuzzy sphere\cite{Mad} and
fuzzy disc\cite{Lizzi:2003ru}. In these examples derivatives are
inner, i.e. they are obtained with the adjoint action of the
generators, and integration is defined with the usual trace. Of interest
here will be the noncommutative plane which is generated by creation
and annihilation operators, ${\bf a}^\dagger$ and ${\bf a}$,
respectively
 \be [{\bf a}, {\bf a}^\dagger]=\BI\;, \label{hocr}\ee $\BI$ being the
 unit operator, with
   derivatives of any function $\Phi$ of  ${\bf a}^\dagger$ and ${\bf
a}$  given by
\be \bar\nabla\Phi=[{\bf a}\;,\Phi]\qquad
 \nabla\Phi=[\Phi,{\bf a}^\dagger]\label{bfnbl} \ee  They satisfy
Leibinz rule along with $[\nabla,\bar\nabla]=0$, $\nabla {\bf
a}=\bar\nabla {\bf a}^\dagger =\BI$ and $\bar\nabla {\bf a}=\nabla
{\bf a}^\dagger
=0$.  The action of a
 free massless scalar field $\Phi$ on the noncommutative plane
 can be written as
\beqa  S_0&=&
 \pi\; {\rm Tr}\;\nabla\Phi\;\bar\nabla\Phi \label{ncafsf}\;\eeqa

More generally, if we are given a pair of generators, ${\bf z}$ and
its hermitean conjugate ${\bf z}^\dagger$, satisfying arbitrary
commutation relations
\be [{\bf z},{\bf z}^\dagger]=
\Theta({\bf z},{\bf z}^\dagger)\;\label{crfzzdgr}\;, \ee it is in general unclear
how to define  derivations or integration.\footnote{Derivations can be
  obtained in special cases.\cite{Behr:2003qc}}  If there  exists an algebraic map from
${\bf z}$ and ${\bf
 z}^\dagger$ to   generators ${\bf a}$ and ${\bf a}^\dagger$ of  the
 noncommutative plane, we can   define
derivatives of functions as in (\ref{bfnbl}), and then apply the
inverse map back  to ${\bf z}$ and ${\bf
 z}^\dagger$. This was  proposed previously \cite{Fosco:2004yz}, but
discussion was restricted to nonsingular maps.  It is
the noncommutative analog of mapping from  generalized coordinates on
a two dimensional manifold ${\cal M}$  to the plane.
 In the commutative setting the existence of a
globally defined map to the plane implies a flat geometry, as opposed
to manifolds ${\cal M}$ with nonvanishing curvature where only local maps to
the plane are defined. Since here we would like to allow for the
possibility of recovering the latter in the commutative limit, we
should admit `local' maps in the noncommutative
setting.  Local in noncommutative theories may be defined by a
restriction of the spectrum of the number operator ${\bf n}={\bf
  a}^\dagger {\bf
 a}$.   We note that singular maps need not imply a singular field
theory.  From (\ref{bfnbl}), ${\bf a}$
and ${\bf a}^\dagger$ appear
 in the derivatives of the fields using the commutator, and so only commutators of ${\bf
 a}$ and ${\bf a}^\dagger$ with the fields need be well defined.  This
may be possible to arrange even if ${\bf a}$ and ${\bf a}^\dagger$
 are not well defined functions of ${\bf z}$ and ${\bf z}^\dagger$ by
 imposing suitable boundary conditions on the fields.  This  was done
 previously in \cite{Pinzul:2001my}.

As is well known, the operator algebra can be realized  on the
(commutative) plane with the use of a star product.  For this,
operators are  mapped to functions (`symbols') on the plane, and
the product of two operators is mapped to the star product of symbols.
Actions such as (\ref{ncafsf}) can then be re-expressed as  integrals on the
plane. Here instead of writing in terms of   symbols of the generators ${\bf a}$ and ${\bf
a}^\dagger$ of the noncommutative plane, we write the
result in terms of
symbols of the operators ${\bf z}$ and ${\bf z}^\dagger$ appearing in (\ref{crfzzdgr}). In the commutative
limit, the latter
symbols reduce to   coordinates, denote them by $x^\mu$, that can be used to
parametrize a coordinate patch of some manifold
${\cal M}$. By performing a derivative expansion, which is equivalent to an
expansion in the noncommutativity parameter, we can then obtain noncommutative
corrections to the commutative action.  In the case of the
 free scalar field $\phi(x)$ the commutative action on any coordinate
 patch ${\cal P}$ is
\be {\cal S}_0^0=\frac 12 \int_{\cal P} d^2x \sqrt{g}\;
g^{\mu\nu}\frac{\partial\phi}{\partial
  x^\mu}\frac{\partial\phi}{\partial x^\nu}\;,\label{aitox} \ee $g$
and $g^{\mu\nu}$ being the determinant and  inverse components,
respectively, of the metric $[g_{\mu\nu}]$. The procedure for
obtaining noncommutative corrections depends on the choice of  star product.  Star products are equivalent if they are
related by a Kontsevich map\cite{Kon}, and so the dynamics computed to
all orders should be identical for equivalent star products.  Common choices for the star
product on the noncommutative plane are the Moyal-Weyl and the Voros. The Voros star product is
based on coherent states which diagonalize ${\bf a}$. An alternative
star product was developed in \cite{Alexanian:2000uz}. It uses an
overcomplete basis of states developed in \cite{mmsz} which instead
diagonalize ${\bf z}$. The Voros star product and the one in
\cite{Alexanian:2000uz} have certain advantages and disadvantages.  
We shall use both of the star products.

 As an example we consider the fuzzy sphere. In
\cite{Alexanian:2000uz} we wrote down a noncommutative analogue of the
stereographic projection of a sphere in terms of  operators ${\bf z}$
and ${\bf z}^\dagger$ satisfying (\ref{crfzzdgr}) for some
$\Theta({\bf z},{\bf z}^\dagger)$.  The noncommutative stereographic
projection is nonsingular, although the map to the noncommutative
plane is not.  For the latter it is necessary to define
 a truncated harmonic oscillator Hilbert
space.  Consequently, we must modify the definition of coherent
states and the star products in this case. 

We introduce the general formalism in  section 2, and discuss the
stereographically projected fuzzy sphere in section 3.  In section 4
we conclude by mentioning possible generalizations of this work.

\section{General Framework}
\setcounter{equation}{0}

We first review free scalar field theory written on a local coordinate
patch in two dimensions.  We include a Poisson structure and a map to the plane. We then consider the
noncommutative version of the theory and apply the above procedure to
compute lowest order corrections.

\subsection{Commutative Theory}

Say    ${\cal P}$ is a coordinate patch for some two dimensional manifold
${\cal M}$.  On  ${\cal P}$ denote the  metric  by
$g_{\mu\nu}(x)$, $x^{\mu}$, $\mu=1,2$, being the coordinates. 
Upon introducing zweibein fields
$e_{\;\;\mu}^a(x)$, $a=1,2$ being the flat index,  the metric can be
expressed by
\be g_{\mu\nu}(x) = e_{\;\;\mu}^a(x) \;e_{\;\;\nu}^a(x)
\ee   The zweibein fields transform vectors to a local orthogonal frame ${\cal O}$, the latter  associated with the  metric $\delta_{ab}$.  If
 $\{\frac{\partial}{\partial x^\mu}\}$ and  $\{\frac{\partial}{\partial y^a}\}
$ are the set of  basis vectors  of  ${\cal P}$ and ${\cal O}$, respectively,
 then
\be \frac{\partial}{\partial x^\mu} = e_{\;\;\mu}^a(x)\;
\frac{\partial}{\partial y^a}\ee  Alternatively, the components
 $v^\mu$ and $u^a$ of tangent  vector $V$ written in the two bases,
 respectively,  are related by $ u^a = e_{\;\;\mu}^a(x)\;v^\mu$.
The set of all orthogonal frames $\{{\cal O}\}$ are related by local
orthogonal transformations.  So if $u'^a$ are components of $V$ in
orthogonal frame  ${\cal O}'$  then \be  u'^a =
\lambda^a_{\;\;b}(x) u^b \;,\label{lotrnsfmtn}\ee  $\lambda(x)= [\lambda^a_{\;\;b}(x)]$
being an orthogonal matrix.  In the tetrad formalism one normally also
 introduces the spin connection.  However here since we shall
only be interested in scalar field theory we do not require
this additional structure.

 Next  we assume a nonsingular Poisson structure on  ${\cal P}$.   So we define a Poisson tensor with components
$\theta^{\mu\nu}(x)=-\theta^{\nu\mu}(x)$.   The  Poisson
 bracket of any two functions $f$ and $g$  on  ${\cal P}$ is
\be\{f,g\}(x)=\theta^{\mu\nu}(x)\;\frac{\partial f}{\partial
x^\mu}\frac{\partial g}{\partial x^\nu}\;\ee  So
\be\{x^\mu,x^\nu\}(x)=\theta^{\mu\nu}(x)\label{pbxmxn}\ee  This is  analogous to the commutation relations (\ref{crfzzdgr}).
Now choose
\be \theta^{\mu\nu}(x)=\theta(x)\epsilon^{\mu\nu}\;,\qquad \theta(x)=\frac1{\sqrt{g(x)}} \label{thtxegi}\ee Then
in  local orthogonal frame ${\cal O}$ the Poisson tensor is mapped to the constant
antisymmetric tensor
 \be \tilde\theta^{ab}=\theta^{\mu\nu}(x)  e_{\;\;\mu}^a(x)
e_{\;\;\nu}^b(x)=\epsilon^{ab}\label{titott}\label{cptnfrmo}\ee   This is true
for the set of  all locally orthogonal frames $\{{\cal O}\}$ (in two dimensions) since the constant  tensor
 is    preserved under  local orthogonal transformations
(\ref{lotrnsfmtn}).  (This  is not the case  in higher dimensions.)
Finally, in order that  $ {\cal O}$  is a  noncommutative plane at lowest order, 
 we need to  impose
 that it defines  a coordinate bases.  Thus the map from ${\cal P}$ to a ${\cal O}$
 should be a coordinate map.  This is since in the  noncommutative
 theory we need an explicit map between the noncommutative analogues
 of the coordinates $x^\mu$ on  ${\cal P}$ and the coordinates
 $y^a$ in frame ${\cal O}$.  The map is in general only valid on
some open region $\sigma$ on  ${\cal O}$.  So $ e_{\;\;\mu}^a(x)=\frac{\partial
 y^a}{\partial x^\mu}$ and as a result \be \{ y^a, y^b\}=\epsilon^{ab}\;,
 \label{loncp}\ee which is the analogue of (\ref{hocr}).

 The basis vectors of ${\cal O}$ can be expressed in terms of the
 Poisson bracket
\be \frac{\partial}{\partial  y^a} =
[\tilde\theta^{-1}]_{ab}\;\{ y^b\;,\;\;\}\;,\ee in analogy to the inner derivatives (\ref{bfnbl}), and then so can
 the basis vectors of  ${\cal P}$
   \be \frac{\partial}{\partial x^\mu}
=- e_{\;\;\mu}^a(x)\;
\epsilon_{ab}\;\{ y^b\;,\;\;\} \label{tddffdrv} \ee  We can apply this
to the case of a scalar field theory  on  ${\cal M}$.  If $\phi$ is a
real  function of coordinates $y$ of ${\cal O}$, then the standard free action on the region $\sigma$ on  ${\cal O}$ is given by
\be {\cal S}_0^0=\frac 1{2}\int_\sigma d^2y \;\{ y^a\;,\phi\}\{
y^a\;,\phi\}\;\label{losftncp}\ee 
  Upon doing  a
  change of variables one gets (\ref{aitox}) on the corresponding
 region $\sigma'$ of  ${\cal P}$.  The action (\ref{aitox}) is exact  for fields with nonvanishing support only on $\sigma'$.

\subsection{Noncommutative analogue}

The prescription to go to the noncommutative plane is to replace real functions by hermitean operators, Poisson
 brackets with $-i\theta_0$ times the commutator, and the integration $\int
 d^2y$ by $2\pi\theta_0\;{\rm Tr}$.  $\theta_0$ is the noncommutativity parameter.
So if we replace $y^a$ by  hermitean
 operators $Y^a$, they  satisfy the Heisenberg algebra
\be [Y^a,Y^b]=i\epsilon^{ab}\theta_0\BI \;, \label{hsnbrg}\ee  which defines the noncommutative plane.  Alternatively,
we can  define creation and annihilation operators ${\bf a}^\dagger$
and ${\bf a}$, respectively
\be {\bf a}=\frac{Y^1+iY^2}{\sqrt{2\theta_0}}\;,\qquad
 {\bf
 a}^\dagger=\frac{Y^1-iY^2}{\sqrt{2\theta_0}}\;,\label{bfaioya}\ee
 satisfying (\ref{hocr}).
  Next introduce the analogue $\Phi$ of the scalar field $\phi$.  It
 is a hermitean function  on the noncommutative plane defined in the
 enveloping algebra generated by $Y^a$.   The free field action   (\ref{losftncp}) goes to
\beqa  S_0&=&-\frac \pi{\theta_0}\; {\rm Tr}_\Sigma\;[ Y^a\;,\Phi]\;[Y^a\;\Phi]\;,\label{SzrtS}
\eeqa or equivalently, (\ref{ncafsf}).    The trace in (\ref{SzrtS}) over all fields $\Phi$ would imply that the dynamics is on the noncommutative plane.  However,  we would like to allow for dynamics on other two dimensional 
noncommutative manifolds $M$.   For this reason we inserted the subscript $\Sigma$ on the trace, which indicates that the expression (\ref{SzrtS}) is
valid for a  restricted set of fields $\Phi$, and it is analogous to the restriction
 in (\ref{losftncp}).  The latter is for fields $\phi$ to have nonvanishing support only on some small region $\sigma$, which we can assume is centered around the origin of coordinate system $\{y^a\}$.  Only in the case of a flat geometry can we take $\sigma$ to be all of $R^2$.  Just as (\ref{losftncp}) is not  `globally' valid for an arbitrary (commutative) manifold ${\cal M}$, the expression (\ref{SzrtS}) cannot be `globally' valid for an arbitrary 
noncommutative manifold $M$.    We should then restrict in
(\ref{SzrtS}) to fields which are defined in a small region   $\Sigma$ of the
noncommutative plane.  By this we mean $\Phi$ acts nontrivially only on eigenstates $|n>$ of the number operator 
${\bf n}={\bf a}^\dagger{\bf a}$ with eigenvalues $n$ sufficiently small  $0\le n<n_0$.  
Say $S$ is the exact expression for the scalar field on $M$.  Then (\ref{SzrtS}) should be a reasonable approximation of $S$ for the case where $\Phi$  vanishes outside $\Sigma$.   This will be 
demonstrated for the fuzzy sphere in the next section.
For fuzzy spaces the Hilbert space is, by definition, finite
dimensional. Moreover, for large dimension $N$ of the Hilbert space, $\theta_0$ is
inversely related to $N$, so that the commutative limit
$\theta_0\rightarrow 0$ corresponds to $N\rightarrow\infty$.  In that
case we can define small as $n_0<<N$.   The approximation is improved by including higher order corrections from $S$.  We assume that the  next order corrections $S_1$ to (\ref{SzrtS}) go like
$\theta_0$, or equivalently $1/N$.  This  is the case for the fuzzy
sphere.   More generally, we  expand the exact expression $S$ for the
action according to
\be S= S_0+S_1+S_2+...\;, \label{expfS}\ee where $S_{m+1}/S_m$ goes like
$\theta_0$.

Instead of starting with (\ref{hsnbrg}), we can examine the
 more general algebra generated by 
 ${\bf z}$ and its hermitean conjugate ${\bf z}^\dagger$, with
commutator given in (\ref{crfzzdgr}) for some  arbitrary function $\Theta({\bf z},{\bf
z}^\dagger)$. Say that the lowest order of the parameter
 $\theta_0$ in
$\Theta({\bf z},{\bf z}^\dagger)$ is linear, and in the commutative  limit $\theta_0\rightarrow 0$, the symbol of
 $\Theta({\bf z},{\bf z}^\dagger)$ tends to $\theta_0\theta(x)$,
where  $\theta(x)$ was given in
 (\ref{thtxegi}).   So (\ref{crfzzdgr}) is the
 noncommutative analogue of  (\ref{pbxmxn}), with  ${\bf z}$ and  ${\bf
 z}^\dagger$  the noncommutative analogues of $x_1+ix_2$ and
 $x_1-ix_2$, respectively, and we recover  the coordinate patch ${\cal P}$
 in the commutative limit.  We already assumed the existence of a
 coordinate map from  ${\cal P}$ to an  orthogonal frame
 ${\cal O}$, and so now we assume that a noncommutative analogue of this is true;
 i.e. that there is an algebraic map from   ${\bf z}$ and ${\bf
 z}^\dagger$ to $Y^a$ (or, equivalently,  ${\bf a}$ and ${\bf
 a}^\dagger$).  The map is in general only `local', meaning that it is only defined  on eigenstates $|n>$ of the number operator 
${\bf n}={\bf a}^\dagger{\bf a}$ with eigenvalues $n$ sufficiently small  $0\le n<n_0$.   The map allows us to re-express the action $S$ `locally' in terms of 
 ${\bf z}$ and ${\bf
 z}^\dagger$.  Next one can introduce 
 a star product, replacing  product of two operators by the star product of their corresponding symbols, and the trace  by integration with respect to some
measure.  Then  the  expansion (\ref{expfS}) maps to a corresponding
expansion of integrals on the plane \be {\cal S}= {\cal S}_0+{\cal
S}_1+{\cal S}_2+...\; \ee We shall express  ${\cal S}_m$  as integrals
in the covariant symbols $\zeta$ and $\bar\zeta$ of ${\bf z}$ and  $
{\bf z}^\dagger$, respectively. There is then a further $\theta_0$ expansion that one can do since  the star product contains
$\theta_0$ to all orders.  So 
\be {\cal S}_m
= {\cal S}_m^0+{\cal S}_m^1 +{\cal S}_m^2+...\;,\ee where
$S_{m}^{p+1}/S_m^p$ goes like $\theta_0$. The zeroth order term ${\cal S}_m^0$ in
the expansion is of order $\theta_0^m$.  It is associated with the ordinary product, and so  it is
star product independent.   In particular, ${\cal S}_0^0$ is the
commutative result, i.e. (\ref{aitox}).  More generally, ${\cal S}_m^p$  is of order $\theta_0^{m+p}$.  As we shall only  be interested in the leading
$\theta_0$ corrections we will only compute ${\cal S}_1^0$ and ${\cal
S}_0^1$, the former being star-product independent.  To compute it we will need the exact expression for the action (or Laplacian).  For the latter we must specify the star-product.  Below we shall compute  ${\cal
S}_0^1$ for the
case of  two different star products: 1) The Voros star product  and
2) the generalized star product of ref. \cite{Alexanian:2000uz}.

\subsection{Voros star product}

  The Voros product is based on standard coherent states on the complex plane. Denote them by
${|\alpha >}_V$,  $\alpha$ being the coordinate on the complex plane, satisfying ${}_V{<\alpha |}  {\alpha
>}_V=1$ and \be{\bf  a}{|\alpha >}_V =\alpha
{|\alpha >}_V \; \ee   The covariant symbol
${\cal A}_V(\alpha,\bar \alpha
)$
of an operator $A$ is a function on the complex plane given by the
matrix element  ${\cal A}_V(\alpha,\bar \alpha
)={}_V{<\alpha |} A {|\alpha
>}_V$. The star product $\star_V$  between
any two covariant symbols ${\cal A}_V(\alpha,\bar \alpha
)$ and ${\cal B}_V(\alpha,\bar \alpha )$  associated with  operators $A$ and
$B$ is defined to be the covariant symbol of the product of operators:
 $$ [{\cal A}_V  \star_V {\cal B}_V](\alpha,\bar \alpha )
\;={}_V{<\alpha |} AB{|\alpha >}_V\; $$ and here
$$\star_V=\exp{\; \;\overleftarrow{ \frac\partial{ \partial\alpha }}\;\;
\overrightarrow{ {\partial\over{ \partial\bar\alpha} }}}$$
Derivatives in $\alpha$ and $\bar \alpha$ are given by \beqa
\frac\partial{ \partial\alpha }\;\; {\cal A}_V(\alpha,\bar \alpha
)&=&{}_V{<\alpha |}[A,{\bf a}^\dagger] {|\alpha
>}_V\cr
& &\cr  {\partial\over{ \partial\bar\alpha}}\;\; {\cal
A}_V(\alpha,\bar \alpha )&=& {}_V{<\alpha |} [{\bf a},A] 
{|\alpha
>}_V\eeqa Defining  $\phi_V$ to be the covariant symbol of the field operator
$\Phi$, the action (\ref{ncafsf}) can  be mapped to \be  {\cal S}_0= \pi\;\int d
\mu_V(\alpha,\bar\alpha)\;\biggl[\frac{\partial\phi_V}{ \partial\alpha
\; }\;\star_V\; {{\partial\phi_V}\over{
\partial\bar\alpha}}\biggr](\alpha,\bar \alpha
)\;,\label{actnnvsrp}\ee where $d \mu_V(\alpha,\bar\alpha)$ is the measure satisfying
the partition of unity $\int d \mu_V(\alpha,\bar\alpha)\;{|\alpha
>_V}\;\;{}_V{<\alpha |}=\BI$.  For the standard coherent states it is
  \be d\mu_V (\alpha,\bar\alpha )\;   =\frac i{2\pi} d\alpha
  \wedge d\bar\alpha\;
\label{imscs}\ee  In the commutative limit
$\theta_0\rightarrow 0$, or $\alpha,\bar\alpha\rightarrow\infty$, and  the
Lagrangian in (\ref{actnnvsrp})  reduces to that of the free scalar
field  (\ref{losftncp}) written in an orthogonal frame $
{\cal O}$.  For this use $y^1 =\sqrt{ \frac {\theta_0}2 }\; (\bar\alpha  +
\alpha)$ and $y^2 =i\sqrt{ \frac {\theta_0}2 }\; (\bar\alpha  -
\alpha)$, which are the covariant symbols of the generators $Y^1$ and
$Y^2$ of the noncommutative plane.

Finally we wish to map back to the coordinate patch ${\cal P}$.
We take it to be spanned by   the covariant symbols  ${\zeta}$ and $\bar{ \zeta}$ of   ${\bf z}$ and  ${\bf z}^\dagger$, respectively, as they have the correct commutative limit.  They are local functions
 of the covariant symbols $\alpha$ and $\bar\alpha$ of ${\bf a}$ and
 ${\bf a}^\dagger$. These functions must then be inverted to express
 the system in terms of  $\zeta$ and ${\bar \zeta}$.  To compare tangent vectors in 
  ${\cal P}$ and ${\cal O}$ one can define the analogue of an
 inverse zweibein matrix
\be h_V = \pmatrix{  h^\zeta_{\;\alpha } & h^\zeta_{\;\bar
  \alpha}\cr  h^{\bar \zeta}_{\;\alpha }& h^{\bar \zeta}_{\;\bar
  \alpha} \cr} =
\pmatrix{\frac{\partial \zeta}{\partial\alpha}&\frac{\partial \zeta}{\partial\bar\alpha}\cr
 \frac{\partial \bar \zeta}{\partial\alpha}&\frac{\partial \bar
 \zeta}{\partial\bar\alpha}\cr}\;,\label{invszbnmtrx}\ee  which goes
like $\sqrt{\theta_0}$ in the commutative limit.   So on 
${\cal P}$ the free scalar field is $\phi(\zeta,\bar\zeta)=\phi_V(\alpha(\zeta,\bar \zeta),\bar
\alpha(\zeta,\bar \zeta) )$, and the  action (\ref{actnnvsrp}) becomes
$$ {\cal S}_0=\frac i2
 \int\frac{  d\zeta
  \wedge d\bar \zeta}{\det h_V}\;{\cal L}_0\;,$$ \beqa  {\cal
 L}_0&=&\sum_{n=0}^\infty\;\frac1{n!}\;\biggl[\biggl(
 h^\zeta_{\;\alpha }\frac{\partial}{ \partial \zeta}+ h^{\bar
 \zeta}_{\;\alpha }\frac{\partial}{ \partial \bar
 \zeta}\biggr)^{n+1}\phi\;\biggr]\;\biggl[\biggl(
 h^\zeta_{\;\bar\alpha }\frac{\partial}{ \partial \zeta}+ h^{\bar
 \zeta}_{\;\bar\alpha }\frac{\partial}{ \partial \bar
 \zeta}\biggr)^{n+1}\phi \biggr] \label{expfvp} \eeqa    The measure, as
 well as the terms in the sum, in general contain different powers of
 the noncommutativity parameter $\theta_0$.  To write the result as an
 expansion in $\theta_0$ one has to expand the components of $h_V$.
  At lowest order in $\theta_0$ we recover the commutative result (\ref{aitox}),
  i.e.  $$ {\cal S}_0^0=\frac i2 \int{  d\zeta \wedge d\bar
  \zeta}\;\sqrt{g}\;{\cal L}_0^0\;,$$
\be {\cal L}_0^0=  g^{\zeta \zeta}\; (\partial\phi)^2 + g^{\bar \zeta\bar \zeta}\;
(\bar\partial\phi)^2 +2 g^{\zeta\bar \zeta}\;
|\partial\phi|^2\;\label{mnsintgd} \ee   Here we set
$\partial=\partial/\partial \zeta$ and $\bar\partial=\partial/\partial
\bar \zeta$ and \be  \pmatrix{  g^{\zeta \zeta} & g^{\zeta \bar \zeta}\cr
g^{\bar \zeta\zeta}&
 g^{\bar \zeta
\bar \zeta}\cr}=\pmatrix{ h^\zeta_{\;\alpha }  h^\zeta_{\;\bar\alpha}&\frac12(
h^\zeta_{\;\alpha }  h^{\bar \zeta}_{\;\bar\alpha} + h^{\bar
\zeta}_{\;\alpha } h^\zeta_{\;\bar\alpha})\cr \frac12( h^\zeta_{\;\alpha } h^{\bar
\zeta}_{\;\bar\alpha} + h^{\bar \zeta}_{\;\alpha } h^\zeta_{\;\bar\alpha}) &
h^{\bar \zeta}_{\;\alpha }  h^{\bar \zeta}_{\;\bar\alpha}\cr
 }\;\ee

\subsection{Generalized star product}

The advantage of the Voros star product is that it has a simple closed
form expression.  On the other hand, the above procedure was complicated by the fact that we had to invert the functions $\zeta(\alpha,\bar \alpha)$ and  $\bar\zeta(\alpha,\bar \alpha)$  in
 order to do the change of variables in (\ref{actnnvsrp}).
This complication is avoided if one can  start with a  star product based on an
overcomplete set of states $\{|\zeta>\}$ which diagonalize ${\bf
z}$ rather than ${\bf a}$\footnote{ For this it is
 necessary that the algebra (\ref{crfzzdgr}) have only  infinite
 dimensional representations.}
   \be {\bf  z}|\zeta>=\zeta|\zeta>\;, \label{zee}\ee $\zeta$ denoting a complex variable.
 The states $\{|\zeta>\}$ were found in
\cite{mmsz} and are a nonlinear deformations of standard coherent states
$\{\widetilde{|\alpha>}\}$ on the complex plane.   The covariant symbol
of  operator $A$, $B$,... are given by  ${\cal A}(\zeta,\bar\zeta)=<\zeta|A|\zeta>$, ${\cal B}(\zeta,\bar\zeta)=<\zeta|B|\zeta>$,... , and their star product by  
$ [ {\cal A} \star {\cal B}](\zeta,\bar \zeta ))=<\zeta|AB|\zeta>$.  From
 $<\zeta|\zeta>=1$ it follows that the complex coordinates $\zeta$
and its complex conjugate $\bar\zeta$ are the
symbols of ${\bf z}$ and ${\bf z}^\dagger$.  The action can then be
directly written in terms of functions of these covariant symbols. 

  The
disadvantage of this approach is that the expression for the
 star product is not  simple unlike the case of the Voros star product. The expression  was obtained in
 \cite{Alexanian:2000uz}. 
 Using the property that the ratio
 $<\zeta|A|\eta>/ <\zeta|\eta>$ is analytic in $\eta$ and
 anti-analytic in $\zeta$, one gets
 \be [ {\cal A} \star {\cal B}](\zeta,\bar \zeta ) = {\cal A}
(\zeta,\bar \zeta )\;\;\int d\mu (\eta,\bar\eta )\;   \;
:\exp{ \overleftarrow{ \frac\partial{ \partial\zeta }}(\eta-\zeta)  }:\;
\; |
<\zeta |\eta >  |^2\;  :
\exp{  (\bar\eta-\bar\zeta ) \overrightarrow{ \frac\partial {
      \partial\bar\zeta } } }:\;\;  {\cal B}(\zeta,\bar \zeta
 )\;,\label{iefsp}\ee   where $ d\mu (\zeta,\bar\zeta )  $ is the
 appropriate measure
  on the complex plane satisfying  the
partition of unity $\int d \mu(\zeta,\bar\zeta)\;{|\zeta>}{<\zeta|}=\BI$.   The colons in (\ref{iefsp}) denote an ordered
  exponential, with the derivatives  ordered to the right in each
  term in the Taylor expansion of $  \exp{  (\eta-\zeta )
    \overrightarrow{ \frac\partial { \partial\zeta } } } \;,$ and to
  the left in each term in the Taylor expansion of $\exp{
    \overleftarrow{ \frac\partial{ \partial\zeta }}(\eta-\zeta)  }$.
Thus
\beqa  :\exp{  (\bar\eta-\bar\zeta ) \overrightarrow{ \frac\partial {
       \partial\bar\zeta } } }: \;&=& 1 + (\bar\eta-\bar\zeta ) \overrightarrow{ \frac\partial {
       \partial\bar\zeta } } +\frac 12  (\bar\eta-\bar\zeta )^2 \overrightarrow{ \frac{\partial^2} {
       \partial\bar\zeta^2 } } +\cdot\cdot\cdot
\cr
:\exp{ \overleftarrow{ \frac\partial{ \partial\zeta }}(\eta-\zeta)
}: \;&=&1 + \overleftarrow{ \frac\partial{ \partial\zeta }}(\eta-\zeta)
+\frac12  \overleftarrow{ \frac{\partial^2}{ \partial\zeta^2 }}(\eta-\zeta)^2
 +\cdot\cdot\cdot\label{eooe} \eeqa The commutative limit is obtained
by performing a derivative expansion, which was done in
\cite{Pinzul:2001my}.  One obtains the following leading three terms
acting on functions of $\zeta$ and $\bar\zeta$:
\be\star\;\; =\;\;  1\;\;+ \;\;\overleftarrow{ \frac\partial{ \partial\zeta }}\; \theta_S(\zeta,\bar\zeta) \;
\overrightarrow{ {\partial\over{ \partial\bar\zeta} }}\;\;
+\;\;
\frac14\biggl[\overleftarrow{\frac{\partial^2}{ \partial\zeta^2 }}
    \;\;  \overrightarrow{ \frac\partial {
      \partial\bar\zeta } }\theta_S(\zeta,\bar\zeta)^2
  \overrightarrow{ \frac\partial {
      \partial\bar\zeta } }\; +\;
\overleftarrow{\frac{\partial}{ \partial\zeta }}
  \theta_S(\zeta,\bar\zeta)^2\;\overleftarrow{\frac{\partial}{ \partial\zeta }}
   \;\;
  \overrightarrow{ \frac{\partial^2} {
      \partial\bar\zeta^2 } }\biggr]\;\;+\;\;\cdot\cdot\cdot
\label{afsvt}\ee
where  $\theta_S(\zeta,\bar\zeta)$ is the  symbol of $
\Theta({{\bf z},{\bf z}^\dagger})  $. [The $S$ subscript distinguishes it from the classical value.]  At lowest order, $\theta_S(\zeta,\bar\zeta)\rightarrow \theta_0 \;\theta(\zeta,\bar\zeta)$.  Since the limit is linear
in $\theta_0$, the derivative expansion is also an expansion in
$\theta_0$.  The  Poisson bracket is recovered from the star
commutator  at leading order.  [The derivation of (\ref{afsvt})
requires  (\ref{zee}).  The latter is not true for the case  of the
fuzzy sphere, as we shall see in the next section.  Thus   (\ref{afsvt}) gets modified for that case.]

Now we return to the free scalar field  with action (\ref{SzrtS}).
We define 
symbol of the field $\Phi$ as
${\phi}(\zeta,\bar\zeta)=<\zeta|\Phi|\zeta>$. In oder to compute its
derivatives we define the  symbols $y^a(\zeta,\bar\zeta)=$  $<\zeta|Y^a|\zeta>$  of $Y^a$, 
$<\zeta|{\bf a}|\zeta> $ of ${\bf a} $  and $<\zeta|{\bf
  a}^\dagger|\zeta> $ of  ${\bf a}^\dagger $.  [The latter two are, in general, not
the same as $\alpha(\zeta,\bar \zeta)$ and $\bar
\alpha(\zeta,\bar \zeta)$ computed previously by taking the inverse of the covariant
symbols
of ${\bf z}$ and ${\bf z}^\dagger$ with respect to the standard
coherent states $|\alpha>_V$.]  Now the action can be expressed as
\be {\cal S}_0'=
\pi{\theta_0}\;
\int_\sigma d \mu(\zeta,\bar\zeta)\;{\cal L}_0'\;,\;\qquad {\cal L}_0'
= -\frac 1{\theta_0^2}({y}^a\star\phi-\phi\star
{y}^a)^{\star 2} \;,\label{Lusgsp}\ee   where   ${\cal A}^{\star 2}=
{\cal A}\star{\cal A}$ and we use the prime to distinguish this result from the one in the previous subsection.  (\ref{Lusgsp}) reduces to (\ref{aitox}) in the
commutative limit $\theta_0\rightarrow 0$; i.e., if we
expand ${\cal L}_{0}'$ in $\theta_0$ the zeroth order term  ${\cal
L}_0^0$ is again given by (\ref{mnsintgd}).  Then in comparing with
(\ref{aitox}),
\be  2\pi{\theta_0}\;
 d \mu(\zeta,\bar\zeta) \rightarrow  i \sqrt{g}\;
 d\zeta\wedge d\bar\zeta\;,\qquad{\rm as} \;\; \theta_0\rightarrow
0\label{clofmsr}  \ee
Going beyond the lowest order,  we have to expand the Lagrangian density $${\cal L}_0'
={\cal L}_0^0+ {\cal L}_0^{1 '}+{\cal L}_0^{2 '}+...\;,$$ as well as
the measure.
 Like in the previous subsection, the first order correction contains quadratic, cubic and quartic terms in derivatives
of $\zeta$ and $\bar\zeta$ of $\phi (\zeta,\bar\zeta)$, but the
coefficients of these terms can differ from the previous results. The
coefficients  can be expressed in terms of $\theta_S$ and
\be h_a = \frac{\theta_S} {\theta_0} \partial {y}^a \qquad \bar h_a = \frac{\theta_S}{\theta_0}\bar
\partial {y}^a \;,\label{invzwbn} \ee and their derivatives.  At lowest order,
$h_a$ and $\bar h_a$ are inverse zweibein components, and
$\theta_S =- i{\theta_0}\epsilon_{ab}h_a\bar h_b$.
   The quadratic terms are
\be  G^{\zeta\zeta}\; (\partial\phi)^2 + G^{\bar\zeta\bar\zeta}\;
(\bar\partial\phi)^2 + 2G^{\zeta\bar\zeta}\; |\partial\phi|^2 \;,
\ee where up to first order in $\theta_S$
\beqa  G^{\bar\zeta\bar\zeta}&=& \overline{ G^{\zeta\zeta}}=- h_a\star
h_a
-\theta_S \bar\partial \theta_S h_a \partial(\theta^{-1}h_a)\cr & &\cr
2    G^{\zeta\bar\zeta}&=&  h_a\star \bar h_a +\bar  h_a\star
h_a+ \theta_S\partial \theta_S h_a\bar \partial(\theta_S^{-1}\bar h_a)+\theta_S
\bar\partial \theta_S \bar h_a \partial(\theta_S^{-1}h_a)\;\label{cttim}\eeqa   The
lowest order terms correspond to the  components of $g^{-1}$.  We can then interpret $G^{\zeta\zeta}$, $G^{\zeta\bar\zeta}$ and $G^{\bar\zeta\bar\zeta}$ as
corrections to the inverse metric.
 The cubic terms are
\be  G^{\zeta,\zeta\zeta}\; \partial\phi \partial^2\phi
+G^{\bar\zeta,\zeta\zeta}\; \bar\partial\phi \partial^2\phi +
G^{\zeta,\zeta\bar\zeta}\; \partial\phi\partial\bar\partial\phi\;\;+\;\;
{\rm complex}\;\;{\rm conjugate} \;, \ee
where
\beqa  G^{\zeta,\zeta\zeta}&=&- \theta_S \bar\partial(\bar h_a\bar h_a)\cr
& &\cr
 G^{\bar\zeta,\zeta\zeta}&=& \theta_S \bar\partial(h_a\bar h_a)\cr & &\cr
 G^{\zeta,\zeta\bar\zeta}&=& \theta_S h_a\bar \partial \bar h_a -
 \theta_S\bar h_a\partial \bar h_a \label{cfofcb} \eeqa The quartic terms are
\be G^{\zeta\zeta,\bar\zeta\bar\zeta}\;| \partial^2\phi|^2
+G^{\zeta\zeta,\zeta\bar\zeta}\; \partial^2\phi \partial\bar\partial\phi  +
G^{\bar\zeta\bar\zeta,\zeta\bar\zeta}\; \bar\partial^2\phi\partial\bar\partial\phi + G^{\zeta\bar\zeta,\zeta\bar\zeta} \;(\partial\bar\partial\phi)^2 \;, \ee
where
\beqa  G^{\zeta\zeta,\bar\zeta\bar\zeta}&=&
G^{\zeta\bar\zeta,\zeta\bar\zeta} = \theta_S \bar h_a h_a\cr
& &\cr
 G^{\bar\zeta\bar\zeta,\zeta\bar\zeta}&=&
\overline{G^{\zeta\zeta,\zeta\bar\zeta}}=-\theta_S h_a h_a
\label{cfofqt} \eeqa
To complete the expansion in $\theta_0$
 we will
 need  the series expansion of $\theta_S(\zeta,\bar\zeta)$, as well
 as of the inverse zweibein $h_a$ about the commutative answers.
 Another issue is  the measure, which is defined to satisfy the
 partition of unity \be \int d\mu(\zeta,\bar\zeta) |\zeta><\zeta|=\BI
 \ee  For the case where the map from ${\bf a}$ and ${\bf a} ^\dagger$
 to  ${\bf z}$ and ${\bf z} ^\dagger$ is of the form \be {\bf z}=
 f({\bf n}+1)\; {\bf a }\qquad\qquad {\bf z}^\dagger= {\bf a
 }^\dagger\;f({\bf n}+1)\; \;,\label{tofrty}\ee the general form of
 the measure was found in terms of an inverse Mellin
 transformation\cite{mnkprA}, \cite{Alexanian:2000uz}. The result can
 in principal be expanded in $\theta_0$, and at  zeroth order one
 should get (\ref{clofmsr}).  This was demonstrated in
 \cite{Alexanian:2000uz} for the fuzzy sphere.  In the next section we
 give the first order correction to the result.

\section{The stereographically projected fuzzy sphere}
\setcounter{equation}{0}

We first review scalar field theory on the sphere.  Because of the
requirement that  local orthogonal frames  form   coordinate bases we
must work with a nonstandard (i.e., nonconformal) metric.

\subsection{Scalar field theory on the commutative sphere}

First start with lowest order fuzzy sphere.  Set the  radius equal
to one.  In terms of embedding coordinates
$x_i,\;i=1,2,3$, the Poisson brackets are
\be \{x_i,x_j\}=\epsilon_{ijk}x_k\;,\qquad x_1^2+x_2^2+x_3^2 = 1\ee
After stereographically  projecting to the complex plane
\be\zeta =\frac{x_1 -ix_2}{1-x_3}\qquad\bar \zeta =\frac{x_1
  +ix_2}{1-x_3}\;,\ee   the Poisson structure is projected to
\be \{\zeta,\bar\zeta\} =- i\theta(|\zeta|^2)\;,\qquad\theta(|\zeta|^2)=\frac 12 (1+|\zeta|^2)^2 \;,\label{prjtdpb} \ee
 which can
be used to construct the K\"ahler two form for $S^2$.
We can then map to a constant Poisson structure
\be \{\alpha,\bar \alpha\}=-i\;, \ee
using \be
\zeta={ \rho(|\zeta|^2})\;\alpha\qquad\qquad \bar\zeta=
{\rho(|\zeta|^2)}\;\bar \alpha \ee  The solution for $\rho(|\zeta|^2)$
is not unique.   It is
\be\frac {|\zeta|^2}{\rho(|\zeta|^2)^2} = C -\frac2{1+|\zeta|^2}\;\label{classlfrrh} \ee
Positivity of  $\rho(|\zeta|^2)^2$ means that the integration constant
satisfies $C\ge 2$.  Furthermore, requiring the mapping to be
nonsingular for all
 $|\zeta|<\infty$  fixes $C=2$, otherwise $\rho $ vanishes at the
origin.  We shall `quantize' about the solution $C=2$. For this
solution
\be \alpha =\frac{\sqrt{2}\; \zeta}{\sqrt{
    1+|\zeta|^2}} \qquad\qquad \bar \alpha =\frac{\sqrt{2}\;\bar
\zeta}{\sqrt{ 1+|\zeta|^2}} \label{naanzts}  \ee  Now set $y^1=(\bar \alpha
+\alpha)/\sqrt{2}$ and $y^2=i(\bar \alpha -\alpha)/\sqrt{2}$ and
substitute into (\ref{invzwbn}) [replacing  $\theta_S/\theta_0$ by $\theta(|\zeta|^2)$] to
get the lowest order inverse zweibein
\beqa h_1
&=&\;\;\frac 14 {\sqrt{1+|\zeta|^2}}\;\;(2+|\zeta|^2 -\bar\zeta^2)
\cr
& &\cr
h_2&=&-\frac i4 {\sqrt{1+|\zeta|^2}}\;(2+|\zeta|^2 +\bar\zeta^2)
\;,\label{loizwbn}\eeqa
and then the lowest order inverse metric
\be  \pmatrix{  g^{\zeta\zeta} & g^{\zeta \bar\zeta}\cr g^{\bar\zeta\zeta}&
 g^{\bar \zeta
\bar\zeta}\cr}=  \frac14 (1+|\zeta|^2)\;
  \pmatrix{(2+|\zeta|^2)\zeta^2&2+2|\zeta|^2
    +|\zeta|^4\cr 2+2|\zeta|^2
    +|\zeta|^4&(2+|\zeta|^2)\bar\zeta^2\cr}\label{zomtrc} \ee
This does  not correspond to the usual conformal  metric for
the sphere, i.e. \be  \pmatrix{  g^{ww} & g^{w \bar w}\cr g^{\bar w w}&
 g^{\bar  w \bar w}\cr}=
  \pmatrix{0&\frac 12 (1+|w|^2)^2\cr\frac 12(1+|w|^2)^2&0\cr}\;\label{ueficm} \ee
  For that one should map tangent vectors
$(\partial/\partial\zeta,\partial/\partial\bar\zeta)$ to some tangent
vectors $(\partial/\partial w,\partial/\partial\bar w)$
\be \frac\partial {\partial\zeta} =  t^w_{\;\zeta }\; \frac\partial
{\partial w} + t^{\bar w}_{\;\zeta }\; \frac\partial
{\partial \bar w} \qquad\qquad \frac\partial {\partial\bar\zeta} =  t^w_{\;\bar
  \zeta} \; \frac\partial
{\partial w} + t^{\bar w}_{\;\bar \zeta}\;  \frac\partial
{\partial \bar w} \;,\ee where
\be \pmatrix{  t^w_{\;\zeta } & t^w_{\;\bar
  \zeta}\cr  t^{\bar w}_{\;\zeta }& t^{\bar w}_{\;\bar \zeta} \cr} = \frac
 {1+|w|^2}{2(1+|\zeta|^2)^{3/2} }
\pmatrix{2+|\zeta|^2&-\zeta^2\cr -\bar
  \zeta^2&2+|\zeta|^2\cr}\;\ee
However this does not correspond to a coordinate map, i.e.  $\partial
 t^w_{\;\zeta
}/\partial\bar \zeta \ne \partial  t^w_{\;\bar
  \zeta}/\partial \zeta$, and  the inverse metric (\ref{ueficm})
cannot be connected to the local orthogonal frame via a coordinate map (which is needed for the noncommutative generalization).
In terms of the inverse  metric (\ref{zomtrc}) the action for the free scalar field is
\be {\cal S}_0^0=\frac i{8} \int \frac{d\zeta\wedge
 d\bar\zeta}{1+|\zeta|^2}\;
\biggl\{\; 4|\partial \phi|^2 \;+\; (2+
|\zeta|^2)\;(\zeta\partial\phi +\bar\zeta
\bar\partial\phi)^2\;\biggr\}\;,\label{zoactn}\ee
where again $\partial=\partial/{\partial\zeta},
\;\bar\partial=\partial/{\partial\bar\zeta}$.  In what follows we look for the lowest order fuzzy corrections to this action.

\subsection{Fuzzy stereographic projection}

For the fuzzy sphere one promotes the coordinates $x_i$
to operators ${\bf x}_i$'s ,
 satisfying commutation relations:
\be [{\bf x}_i,{\bf x}_j] = {i \beta}\;\epsilon_{ijk}{\bf x}_k\;,\label{xixj}\ee
as well as   $ {\bf x}_i{\bf x}_i=\BI$.  When the parameter $ \beta$ has values ${1\over {
\sqrt{j(j+1)}}}\;,
\; j={1\over 2}, 1, {3\over 2},...$ , ${\bf x}_i$
 has finite dimensional representations,
which are simply given by ${\bf x}_i=\beta {\bf J}_i$,
 ${\bf J}_i$ being the angular momentum matrices.  Denote the $N=2j+1$ states of an irreducible
representation $\Gamma^j$ as usual by $|j,m>$, $m=-j,-j+1,...,j$, spanning  Hilbert space $H^j$.   The commutative
limit is $j\rightarrow\infty$ corresponding to infinite dimensional
representations. 

 The operator analogue of the stereographic projection  to a
pair of operators ${\bf z}$ and ${\bf z}^\dagger$ is defined up to an ordering
ambiguity.
  We  fix it as follows:
\be {\bf z} =({\bf x}_1-i {\bf x}_2)(1-{\bf x}_3 )^{-1} \;,\qquad
   {\bf z}^\dagger =(1-{\bf x}_3 )^{-1} ({\bf x}_1+i {\bf x}_2)
   \label{fzysp} \ee {\it We remark that this transformation is
nonsingular for all finite values of $j$ since the eigenvalues of
${\bf x}_3$ are less than one.}  We can now define the operator
$\Theta({\bf z},{\bf z}^\dagger)$ appearing in (\ref{crfzzdgr}).  It is
diagonal on  $H^j$
$$  \Theta({\bf z},{\bf
  z}^\dagger)|j,m>= \theta^j_m|j,m>\;, $$
\be  \theta^j_m = \frac{j(j+1)-m^2-m}{(\sqrt{j(j+1)}-m-1)^2}-\frac{j(j+1)-m^2+m}{(\sqrt{j(j+1)}-m)^2} \label{egvlsfthta} \ee
The two terms in (\ref{egvlsfthta}) are eigenvalues of ${\bf zz}^\dagger$ and ${\bf z}^\dagger
{\bf z}$, respectively.  An explicit expression for $  \Theta({\bf z},{\bf
  z}^\dagger)$ in terms of just ${\bf zz}^\dagger$ was given in \cite{Alexanian:2000uz}.
It is \be \Theta({\bf z},{\bf
  z}^\dagger) = \beta \chi\; \biggl(1+{\bf zz}^\dagger -{1\over 2}
\chi\; (1+{\beta\over 2}{\bf zz}^\dagger ) \biggr)  \;,\label{zzdag}
\ee where
$${\beta \over 2}  \chi = 1 +{\beta \over {2\xi}}  -\sqrt{ {1\over \xi}+\biggl(
{\beta \over{2\xi}}\biggr)^2}\;,\qquad
\qquad \xi=1+\beta {\bf zz}^\dagger$$
Expanding this result in $1/j$ gives the classical answer plus the
 next order correction
\be\Theta({\bf z},{\bf
  z}^\dagger)\rightarrow \frac1{2j} (1+{\bf zz}^\dagger)^2-
\frac1{4j^2} (1+{\bf zz}^\dagger)^3\;,\qquad {\rm as}\;\;j\rightarrow \infty   \;.\label{lorhs}\ee

We next define the  map from the  harmonic oscillator algebra.
This is clearly a singular map since $H^j$ is finite dimensional and the Hilbert space
 ${\tt H}$ for the latter is not.
  For irreducible representation $\Gamma^j$, we can restrict the map to act on
the finite dimensional
subspace of ${\tt H}$ spanned by the first $2j+1$ eigenstates $|n>  $,
$n=0,1,2,...,2j$, of the number operator ${\bf n}={\bf a}^\dagger{\bf a}$.
More precisely,  we identify
$|j,m>$ in $H^j$ with $|j+m>  $ of ${\tt H}$ , and the  map is applied to this subspace.
  For the map we take the ansatz (\ref{tofrty}).  Because the function
 $f$ depends on $j$ we include a $j$ subscript \be {\bf z}= f_j({\bf
 n}+1)\; {\bf a } \label{mzta}\ee Using \be \Theta({\bf z},{\bf
  z}^\dagger)= ({\bf n}+1)  f_j({\bf  n}+1)^2-{\bf n}  f_j({\bf
  n})^2\;,\ee it follows that
 \be f_j({\bf  n}) = \frac{\sqrt{2j-{\bf  n }+1}}{\sqrt{j(j+1)} + j  - {\bf  n}}
  \label{fjn}  \ee   
$f_j({\bf  n})$ is zero when acting on $|2j+1>$, and hence ${\bf z}^\dagger|2j> =0$.
 Therefore ${\bf z}$ and ${\bf z}^\dagger$ have a well defined action
 on the first $N=2j+1$ harmonic oscillator states, corresponding to the physical
 Hilbert space for the fuzzy sphere,  and are ill-defined on
states with $n>2j+1$. 

On states with  $n\le 2j$, ${\bf z}$ goes like
\be {\bf z}\rightarrow {\bf a}\;\frac 1{\sqrt{2j
-{\bf n}}}\;\biggl(1\;+\;{\cal O} (1/j^2)\;\biggr)\;,\label{oojexpfz}\ee
in the large $j$ limit.
  The inverse of (\ref{mzta}) cannot be well defined since ${\bf
a}^\dagger$ takes vectors out of the physical Hilbert space.  As
pointed out previously, we  only need that the commutators of  ${\bf
a}$ and ${\bf a}^\dagger$ with the fields  be well defined, and this can be arranged by imposing suitable boundary
conditions on the fields. Fields $\Phi$ are defined in the enveloping
algebra generated by ${\bf z}$ and ${\bf z}^\dagger$ and hence are
nonvanishing on the same $N$-dimensional subspace of ${\tt H}$.
In order for the derivatives $\nabla\Phi$ and $\bar\nabla\Phi$ to be
defined on the same subspace we need $<2j|\Phi
=0$  and $\Phi|2j>=0$, respectively.  For the free field action $S_0$ in (\ref{ncafsf})
we then need both of these conditions.  $S_0$ with these  boundary
conditions is then the
 free field action on the fuzzy disc.\cite{Lizzi:2003ru}   It also
serves to approximate the  free field action $S$ on the fuzzy sphere
in the limit $n<<j$ as we show in the next subsection.

 Although the inverse of (\ref{mzta}) is not  well defined, we know
from (\ref{naanzts}) that the  commutative analogue (\ref{naanzts}) in  a local coordinate patch containing the origin.   With this in
mind, we attempt to make an asymptotic expansion of  ${\bf a}$ and
${\bf
  a}^\dagger$ as a function of  ${\bf z}$ and ${\bf z}^\dagger$.
First write
  $ f_j({\bf  n})$  in terms of just ${\bf zz}^\dagger$ using \be {\bf n} = j
+ \sqrt{j(j+1)} -\frac2{\beta\chi} \ee  Upon expanding in $1/j$
\be f_j({\bf  n}+1)^2\rightarrow \frac1{2j} \;(1+{\bf zz}^\dagger) +{\cal
  O}(1/{j^3})\;,\qquad {\rm as}\;\;j\rightarrow \infty
\label{ojexpfrfs}
\ee  Then  from (\ref{mzta})
\be  {\bf a}\rightarrow \frac{\sqrt{2j}}{\sqrt{1+{\bf
      zz}^\dagger}}\;{\bf z} +{\cal
  O}(1/{j^{3/2}})\;,\qquad {\rm as}\;\;j\rightarrow \infty\;,
\label{ojexpfrbda}
\ee  which reveals that we have the fuzzyfied about the $C=2$ solution
(\ref{naanzts}). 

\subsection{Star product independent  correction}

Here we compute ${\cal S}_1^0$.  We start with the standard action for
a field $\Phi$ on fuzzy $S^2$. \be S= -\frac{\pi}{2j+1}\; {\rm Tr}\;
[{\bf J}_i, \Phi]^2  \label {eaofs2} \ee It can be re-expressed on the
truncated harmonic oscillator Hilbert space using the
Holstein-Primakoff map\cite{hp} \be {\bf J}_+ = \sqrt{2j -{\bf n}
+1}\;{\bf a}^\dagger\qquad {\bf J}_- = {\bf a}\;\sqrt{2j -{\bf n}
+1}\qquad {\bf J}_3 = {\bf n} -j \ee  Just as with the raising
operator ${\bf z} ^\dagger$,  ${\bf J}_+ $ annihilates the highest
state,    ${\bf J} _+|2j>=0 $.  Substituting into (\ref{eaofs2}) gives
\be S= -\frac{\pi}{2j+1}\; {\rm Tr}\;\biggl( [{\bf a}^\dagger  \sqrt
{2j -{\bf n}},  \Phi]\;[\sqrt{2j -{\bf n}}\;{\bf a},\Phi]\;+\; [{\bf
n},\Phi]^2\biggr)  \label{hpmoeaofs2} \ee  To recover (\ref{ncafsf}), consider the
case of $\Phi$ vanishing on all states
$|n>$ with $n$ greater than some $n_0<<j$.  (\ref{ncafsf}) appears in the limit of large $j$.
The next order term $S_1$ in the expansion (\ref{expfS}) in $1/j$ is   \beqa
S_1&=&\frac{\pi}{4j}\; {\rm Tr}\;\biggl([{\bf na}^\dagger,\Phi][{\bf
a},\Phi]+[{\bf a}
^\dagger,\Phi][{\bf an},\Phi] - 2[{\bf n},\Phi]^2\biggr)  \cr &
&\cr &=& \frac{\pi}{4j}\; {\rm Tr}\;\biggl(2\nabla\Phi\bar\nabla\Phi
-2 {\bf n} [\nabla\Phi,\bar\nabla\Phi]- ({\bf a} \nabla\Phi)^2-({\bf
a}
^\dagger\bar \nabla\Phi)^2\biggr) \eeqa  This can be expressed as
an integral over the symbols $\alpha$ and $\bar\alpha$  of  ${\bf a}$
and ${\bf a}^\dagger$.  At lowest order in $1/j$ one gets \be {\cal
S}_1^0 = \frac{ i}{4j}\;\int d\alpha\wedge d\bar\alpha \;
\biggl\{  \frac {\partial \phi}{\partial\alpha}\frac{\partial
\phi} {\partial\bar\alpha} - \frac{\alpha^2}2\biggl(\frac{\partial
\phi} {\partial\alpha}\biggr)^2-
\frac{\bar\alpha^2}2\biggl(\frac{\partial
\phi}{\partial\bar\alpha}\biggr)^2\biggl\}\;,\ee where $\Phi$ is
again the symbol of $\Phi$.  This is the correction to the scalar
field action in the local orthogonal frame ${\cal O}$.  Since it is the
lowest order result it is independent of the star product.  Finally, applying
the inverse map back the coordinate patch ${\cal P}$, one gets the following
first order correction to (\ref{zoactn})\be {\cal S}_1^0 = \frac{
i}{16j}\;\int
\;\frac {d\zeta\wedge d\bar\zeta}{(1+|\zeta|^2)^2}\;\biggl\{
4(1+3|\zeta|^2)
|\partial\phi|^2 - (2+|\zeta|^2+|\zeta|^4)
(\zeta\partial\phi +\bar\zeta\bar\partial\phi)^2\;\biggr\}\ee

\subsection{Star product dependent  correction}

We next compute the star product dependent first order correction $
{\cal S}^1_0$ to (\ref{zoactn}).   We  again consider the Voros
star product, along with the generalized star product of \cite{Alexanian:2000uz}.
 However now we need to modify the coherent states by making a finite
truncation of the sum over harmonic oscillator states, and as a result, we
 modify the corresponding star products. The reason for  the
truncation is because ${\bf z}$ and ${\bf z}^\dagger$ are only defined
on the $2j+1$-dimensional subspace of ${\tt H}$.  The resulting
coherent states will no longer be eigenstates of either ${\bf a}$ or
${\bf z}$, although they tend to eigenstates in the commutative limit.
 Our procedure  requires  a star product written
directly on  the coordinate patch, and therefore excludes those such
as \cite{Balachandran:2001dd}  for the sphere which are expressed in terms of embedding coordinates.

\subsubsection{Truncated Voros star product}

In \cite{Pinzul:2001qh} we examined the star product based on a
truncation of the standard coherent states.    Here we truncate at $(2j)^{th}$
excited state of the harmonic oscillator, and denote the resulting
coherent states by $|\alpha ,j>_V$,
\be
|\alpha ,j>_V=   N_{V,j}(|\alpha|^2)^{-\frac12} \;\sum^{2j}_{n=0}
\frac{1}{n !}({\alpha}{\bf a}^\dagger)^n |0>\ \;,
\ee $|0>$ being the harmonic oscillator ground state.  The requirement that
  $|\alpha,j>_V$ are unit vectors fixes $ N_{V,j}(|\alpha|^2)$:
\be  N_{V,j}  (|\alpha|^2)=\sum^{2j}_{n=0}
\frac{1}{n !}|\alpha|^{2n}\equiv
\mbox{e}_{2j}(|\alpha|^2)\ .
\ee
The truncated coherent state is almost (up to the harmonic oscillator state
$|2j>$) an eigenstate of ${\bf a}$
\be
{\bf a} |\alpha,j>_V = \alpha |\alpha,j>_V \;-\;
 \frac { \alpha^{2j+1}} {\sqrt{(2j)!  \; \mbox{e}_{2j} (|\alpha|^2)
 }}
\;   |2j>\
\label{aee}\ee
As a result
$$  {}_V<\alpha,j|{\bf a}|\alpha,j>_{V}  =\alpha\;(1 -\mu_j(|\alpha|^2)\;)
\qquad\qquad {}_V<\alpha,j|{\bf a}^\dagger|\alpha,j>_{V}  =\bar\alpha\;
(1 -\mu_j(|\alpha|^2)\;)$$\be\mu_j(x)= \frac{x^{2j}} {(2j)!
\;e_{2j}(x)}\ee

       Below we shall only consider the lowest order corrections to
the classical result, which are of order $1/j$.  Since $\mu_j(x)$
vanishes much more rapidly than that we are then
 justified in approximating $\alpha$ and $\bar\alpha$ as the covariant
symbols of ${\bf a}$ and ${\bf a}^\dagger$, respectively. Moreover,
the corrections to Voros star product will be negligible.  The exact
expression is obtained from (\ref{iefsp}).  The scalar product
appearing there
can be written as a truncated exponential $e_{2j}$.
For large $j$ (compared to its argument) it rapidly approaches the
scalar product of the usual coherent states.  The same is true for the
integration measure $ d\mu_V (\alpha,\bar\alpha )$.  The exact
expression was computed in
  \cite{Pinzul:2001qh}.  The result is \be d\mu_V (\alpha,\bar\alpha )
= \frac i{2\pi }\Theta_{2j}(|\alpha |^2 ) \;d\alpha \wedge
d\bar\alpha \;,\ee with \be \Theta_N(|\alpha|^2) = \frac{\Gamma
[N+1,|\alpha|^2 ]}{\Gamma[N+1]}= e^{-|\alpha|^2}e_N (|\alpha|^2)
\;,\ee where $\Gamma [N,x ]$ denotes the incomplete gamma function.
For $j>>1$ the difference of $\Theta_{2j}$ and $1$ is exponentially
small.\footnote{Another interesting limit of $\Theta_N$ was found in
\cite{Pinzul:2001qh},\cite{Lizzi:2003ru}.   It is the limit of the
disc. For this one re-scales the coordinates $\alpha$ and $\bar\alpha$
by a factor of $1/\sqrt{\theta_0} $ and sets
 $N\rightarrow \infty  $ and $\theta_0 \rightarrow 1/N$.  Then
 $\Theta_N$ goes to the characteristic function  on a unit disc:
$$\Theta_N(|\alpha|^2/\theta_0 ) \rightarrow
\left\{\matrix{1 \;,\quad |\alpha|<1 \cr 0\;,\quad |\alpha|>1
    \cr}\right. \; $$ }

Next we use (\ref {oojexpfz}) to compute  the covariant symbols of
${\bf z}$ and ${\bf z}^\dagger$ for large $j$.  The result is
\be \zeta(\alpha,\bar\alpha) \rightarrow \frac{\alpha}{\sqrt{2j -|\alpha|^2}}\;
\biggl(1\;+\; \frac{j -\frac18 |\alpha|^2}{(2j -|\alpha|^2)^2}\;+\;{\cal O}(1/j^2)
\;\biggr)\;,\label{smblzna}\ee with  the covariant symbol of
${\bf z}^\dagger$ being the complex conjugate. Since (\ref {oojexpfz})
is only valid for
 $n\le 2j$, (\ref{smblzna}) will be only valid for $|\alpha|^2
\le 2j$. Upon inverting this
expression we get the first order correction to the commutative result
(\ref{naanzts})
\be \alpha(\zeta,\bar \zeta) =
\frac{\sqrt{2j}\;\zeta}{\sqrt{1+|\zeta|^2}}\biggl(1-\frac 1{16j}(4+3|\zeta|^2)+{\cal
  O}(1/{j^{2}})\biggr) \label{vfoctalp}
\ee  The correction is small provided $|\zeta|^2<<j$.
It remains to compute the inverse zweibein matrix (\ref{invszbnmtrx}).
We get
\beqa h^\zeta_{\;\;\alpha} &=& h^{\bar \zeta}_{\;\;\bar \alpha}=
\sqrt{\frac{1+|\zeta|^2}{2j}} \;\biggl\{1 +\frac{|\zeta|^2}2+ \frac\eta{16j}
\
+{\cal O}(1/j^2)\biggl\}\cr& &\cr h^\zeta_{\;\;\bar\alpha}
&=&\overline{h^{\bar \zeta}_{\;\;\alpha}}=\frac{\zeta^2}2\;
\sqrt{\frac{1+|\zeta|^2}{2j}} \;\biggl\{1 + \frac\xi {16j}
+{\cal O}(1/j^2)\biggl\}\;,
\eeqa where \beqa \eta &=& 4+8|\zeta|^2 +\frac{15}2|\zeta|^4+3|\zeta|^6
\cr & &\cr \xi &=& 10+15|\zeta|^2+6|\zeta|^4 \eeqa The result can then be substituted
into (\ref{expfvp}) and the measure to give the  correction ${\cal
S}_0^1$.   Keeping the first order  corrections to the $n=0$
contribution to the sum in (\ref{expfvp}) we get
\beqa |\;( h^\zeta_{\;\alpha
 }{\partial}+ h^{\bar \zeta}_{\;\alpha }{\bar\partial}
)\;\phi\;|^2&\rightarrow & \frac{(1+|\zeta|^2)}{2j}\;\biggl\{
|\partial\phi|^2 +\frac14 (2+|\zeta|^2)(\zeta\partial\phi
+\bar\zeta\bar\partial\phi)^2\\& &\cr& &\qquad
\;+\; \frac1{16j} \biggl( (2\eta-|\zeta|^2\xi) |\partial\phi\;|^2
+\frac14 (2\eta  +2\xi +|\zeta|^2\xi)(\zeta\partial\phi
+\bar\zeta\bar\partial\phi)^2 \biggr)\;\biggr\}\;,\nonumber
\eeqa while   the lowest order correction to the $n=1$ term is
\beqa |\;( h^\zeta_{\;\alpha
 }{\partial}+ h^{\bar \zeta}_{\;\alpha }{\bar\partial}
)^{2}\;\phi\;|^2&\rightarrow & \frac{(1+|\zeta|^2)^2}{64j^2}\;\bigg|\;
3\zeta^3\;\partial\phi
\;+\;(4+3|\zeta|^2)\zeta\;\bar\partial\phi\label{neqlone}\\
& &\cr & &\quad +\;\zeta^4\;\partial^2\phi\;+\;2\zeta^2(2+|\zeta|^2)\;
\partial\bar\partial\phi \;+\;(2+|\zeta|^2)^2\;\bar\partial^2\phi\;\bigg|^2\nonumber\;\eeqa
Up to first order the determinant becomes  \be \det
h_V=\frac1{2j}(1+|\zeta|^2)^2\;\biggl[1+\frac1{16j}\;\frac{(2+|\zeta|^2)\eta
-\frac12 |\zeta|^4\xi }{1+|\zeta|^2}\biggr]
\ee Combining these results gives
\beqa {\cal S}_0^1&= & \frac i{32j}\int d\zeta\wedge d\bar\zeta\;\;\biggl\{
\frac{|\zeta|^2(\eta-\xi) -\frac12|\zeta|^4\xi}{(1+|\zeta|^2)^2}\;|\partial\phi|^2
\\ & &\cr & &\;\; +\; \frac{2(\xi-\eta) + (3\xi-2\eta)\;|\zeta|^2 +(2\xi
-\eta)|\zeta|^4 -\frac12\xi|\zeta|^6}{4(1+|\zeta|^2)^2}\;(\zeta\partial\phi
+\bar\zeta\bar\partial\phi)^2\cr & & \cr & &\;\;+
\;\frac12\;\bigg|\; 3\zeta^3\partial\phi
+(4+3|\zeta|^2)\zeta\bar\partial\phi
+\zeta^4\partial^2\phi+2\zeta^2(2+|\zeta|^2)
\partial\bar\partial\phi +(2+|\zeta|^2)^2\bar\partial^2\phi\;\bigg|^2\;\bigg\}
\nonumber
\eeqa

\subsubsection{Truncated generalized star product}

 We  denoted the truncated generalized coherent states by $|
\zeta,j>$
in \cite{Alexanian:2000uz}.  They were written as a finite series  of
 eigenstates of ${\bf  n}$, and as a result are no longer eigenstates of $
{\bf z}$. Instead,
\be {\bf  z}| \zeta,j> =\zeta |\zeta,j> -\; \frac{N_j (|\zeta |^2  )^{-\frac12}\;
\zeta^{2j+1}} {\sqrt{(2j)!} \;[f_j(2j)]!     }\;|2j>     \;,\label{zeig}\ee
where  $[f(n)]!= f(n)f(n-1)...f(0)$ and $ N_j (x  )$ is a
normalization function.  From the demand that $| \zeta,j>$ has unit
norm,
\be N_j (x )  =
\sum_{n=0}^{2j}
\frac{x^n} {n!\;([f_j(n)]!)^2} \;,\label{Nsj}\ee which can be
expressed in terms of a hypergeometric function \be   N_j (x  )
 =\;\frac {\Gamma(\gamma+2j+ 1)\;{}^2}
{(2j+1)!\;(2j)!\;\Gamma(\gamma)\;{}^2} \;\;
 {}_3F_2 (1,1,-2j;\gamma,\gamma; -x^{-1}) \;x^{2j} \;, \label{ceffj}
\ee
where $\gamma = \sqrt{j(j+1)} - j $.  So now $\zeta$ and $\bar\zeta$ are not
covariant symbols of ${\bf z}$ and ${\bf z}^\dagger$, respectively.  Rather,
$$  <\zeta,j|{\bf z}|\zeta,j>  =\zeta\;(1 -\nu_j(|\zeta|^2)\;)\qquad\qquad
 <\zeta,j|{\bf z}^\dagger|\zeta,j>  =\bar\zeta\;(1
-\nu_j(|\zeta|^2)\;)$$  \be\nu_j(x)= \frac{x^{2j}} {(2j)!
\;([f_j(2j)]!)^2N_j(x)}\ee   Next we  expand in $1/j$.   The
asymptotic behavior of $N_j (x )$ is
\be
N_j(x)\sim
(1+x)^{2j}\left(\frac{2j}{1+x}\right)^{2(1-\gamma)}
\exp{\biggl(\frac{1+x}{8j}\biggr) }\;, \label{cloN}\ee which is
valid for $|\zeta|^2=x\ll j\;.  $  Using ${(2j)!} \;([f_j(2j)]!)^2     \sim
{2\pi j}$, $\nu_j(x)$ goes like
\be \nu_j(x)\sim \frac 1{4\pi j^2}\;\frac {x^{2j}\; e^{-\frac{
      1+x}{8j}}}{(1+x)^{2j-1}}\;\;,\qquad {\rm as}\;\;j\rightarrow
\infty \ee   Below we shall only consider the lowest order corrections
to the classical result, which are of order $1/j$. Therefore the
correction $\nu_j(x)$ can be ignored, and we are
 justified in approximating $\zeta$ and $\bar\zeta$ as the covariant
symbols of ${\bf z}$ and ${\bf z}^\dagger$, respectively. Using these
coherent states we can directly compute the symbols of $\Theta({\bf z},{\bf
  z}^\dagger)$ and ${\bf a}$, which was not the case using the Voros
product.  So from (\ref{lorhs}) the symbol $\theta_S(\zeta,\bar\zeta)$
for $\Theta({\bf z},{\bf
  z}^\dagger)$  can be approximated up to the first order corrections
by \beqa\theta_S(\zeta,\bar\zeta)&\rightarrow & \frac1{2j}
(1+\zeta\star\bar\zeta)^{\star 2}-
\frac1{4j^2} (1+\zeta\star\bar\zeta)^{\star 3}\;,\qquad {\rm as}\;\;j\rightarrow \infty
\eeqa   If we apply the expansion (\ref{afsvt}) for
the star product this leads to
\beqa\theta_S(\zeta,\bar\zeta)&= & \frac1{2j}
(1+|\zeta|^2)^2\;\biggl(1+
\frac1{2j} (1+2|\zeta|^2)+{\cal
  O}(1/{j^{2}})\biggr) \label{focttht}
  \;\eeqa  The lowest order term
corresponds to $\frac 1j\theta(|\zeta|^2)$ in (\ref{prjtdpb}).  If we again
use $\alpha$ to denote the symbol of ${\bf a}$, then from
(\ref{ojexpfrbda}) we get
\be  \alpha =
\sqrt{2j}\;(1+\zeta\star\bar\zeta)^{\star (-\frac 12)}\star \zeta +{\cal
  O}(1/{j^{3/2}})\ee  Then we can expand the definition ${\cal A}^{\star (-\frac
  12)}\star{\cal A}^{\star (-\frac 12)}\star{\cal A} = 1$ to compute
the first order  correction to (\ref{naanzts})
\be \alpha =
\frac{\sqrt{2j}\;\zeta}{\sqrt{1+|\zeta|^2}}\biggl(1-\frac 1{16j}(4+|\zeta|^2)+{\cal
  O}(1/{j^{2}})\biggr) \label{foctalp}\;,
\ee  which differs slightly from the analogous expression (\ref{vfoctalp})
obtained using the Voros star product.   Once again the region of validity  is $|\zeta|^2<<j$.

In the above we have used the star product expansion (\ref{afsvt}) up
to order $1/j$.  However to obtain the leading correction to the
commutative action we must expand ${y}^a\star\phi$ and $\phi\star
{y}^a$ in (\ref{Lusgsp}) up to $1/j^2$ since the difference, or star
commutator, appears there.  Because $\zeta$ and
$\bar\zeta$ are symbols of ${\bf z}$ and ${\bf z}^\dagger$ only up to
order $1/j^2$, there will be corrections to the star product expansion
(\ref{afsvt}) up to this order.  For example, while  the term \be     {\cal A}(\zeta,\bar \zeta )
 \;\;\int d\mu (\eta,\bar\eta )\;   \;
 \overleftarrow{ \frac\partial{ \partial\zeta }}(\eta-\zeta)
\;\; <\zeta,j |\eta,j ><\eta,j |\zeta ,j>\;    \;
\; {\cal B}(\zeta,\bar \zeta ) \;\ee vanishes when  $\zeta$ is the symbol
 of ${\bf z}$, it now produces  $1/j^2$ corrections.  Using (\ref{zeig}) this becomes $$
\frac\partial{ \partial\zeta } {\cal A}(\zeta,\bar \zeta )
 \; {\cal B}(\zeta,\bar \zeta )\; \int d\mu (\eta,\bar\eta )\;
\biggl\{ <\zeta,j|{\bf z} |\eta,j ><\eta ,j|\zeta,j >-<\zeta,j |\eta,j ><\eta ,j|{\bf
z}|\zeta,j
>
$$  $$\qquad \qquad+\;\;\frac1 {\sqrt{(2j)! }\;[f_j(2j)]!
}\biggl(\;\frac{\eta^{2j+1}}{N_j (|\eta |^2  )^{\frac12}}<\zeta,j|2j>
<\eta,j|\zeta,j>$$ \be\qquad\qquad\qquad\qquad -\;\;
\frac{\zeta^{2j+1}}{N_j (|\zeta |^2 )^{\frac12}}
<\zeta,j|\eta,2j><\eta,j|2j>\biggr)\;\biggr\}
\ee
The first line in the braces vanishes by the partition of unity. Using
\be<\eta,j |
\zeta,j
>
=   N_j (|\eta |^2  )^{-\frac12} \; N_j (|\zeta |^2  )^{-\frac12}
 N_j(\bar\eta\zeta) \;,\ee we get \be  \frac\partial{ \partial\zeta }
{\cal A}(\zeta,\bar \zeta )
 \;\; {\cal B}(\zeta,\bar \zeta )\;\biggl[\;\int d\mu (\eta,\bar\eta
 )\;\frac{\bar\zeta^{2j}\eta^{2j+1}\;N_j(\bar\eta\zeta)}{(2j)!
\;([f_j(2j)]!)^2\;N_j(|\zeta|^2)\;N_j(|\eta|^2)}\;\;-\;\;\zeta\nu_j
(|\zeta|^2)\;\biggr]\;,\ee which goes like $1/j^2$.  Analogous
 $1/j^2$ corrections come from
\beqa
  & &  {\cal A}(\zeta,\bar \zeta ) \;\;\int d\mu (\eta,\bar\eta )\; \;
 <\zeta,j |\eta ,j><\eta,j |\zeta ,j>\;    \;
  (\bar\eta-\bar\zeta ) \overrightarrow{ \frac\partial {
 \partial\bar\zeta } } \;\; {\cal B}(\zeta,\bar \zeta )\eeqa So upon
expanding ${y}^a\star\phi$ and $\phi\star {y}^a$ in the Lagrangian
(\ref{Lusgsp}) up to $1/j^2$ we get terms involving single derivatives
of $\phi$, as well as terms proportional to $\phi^2$. Such terms did
not appear using the truncated Voros star product.  There will also be
corrections to the coefficients of the quadratic terms (\ref{cttim}).

Next we write down corrections to  the classical measure. We set
\be  \frac{2\pi}j\; d\mu (\zeta,\bar\zeta )  =
H_j (|\zeta |^2  ) \;d\zeta \wedge d\bar\zeta\;, \ee
corresponding to the left hand side of (\ref{clofmsr}).
 In \cite{Alexanian:2000uz} we found an exact expression for $H_j (x)$
in terms of  hypergeometric function ${}_2F_1$:
\be
H_j(x)=\frac ij\;\;{N_j(x)}\;\;{}_2F_1(\gamma +2j+1,\gamma +2j+1;2j+2;-x),\label{measure}
\ee where again  $\gamma = \sqrt{j(j+1)} - j $ and  $N_j(x)$ was given
in (\ref{Nsj}) and (\ref{ceffj}).   The asymptotic
expansion of ${}_2F_1$ for large parameters \cite{luke} is
\be {}_2F_1 (a_1+2j,a_2+2j;b+2j; -x) \sim
(1+x)^{b-a_1-a_2-2j} \biggl(1\;-\;\frac{(b-a_1)(b-a_2) x}{2j}\;
+\;{\cal
  O}(1/{j^{2}})\biggr) \;,\;\ee   while the   asymptotic form  of
$N_j(x)$ was given in (\ref{cloN}). Both of these expressions are only
valid for $ x\ll j$.  Then\be H_j(x)\sim  \frac{2 i}{
(1+|\zeta|^2)^2}\;\biggl( 1+\frac{1+2\ln{2j}}{8j}+{\cal
  O}(1/{j^{2}})\biggl)\ee

In what remains we compute the coefficients of the cubic and quartic
terms which are unaffected by the above considerations. Call $y^a$ the
symbol for $Y^a$, and set $\theta_0=1/j$. Then from (\ref{bfaioya})
\be \alpha =\sqrt{\frac j2}\;(y^1+iy^2)\;\ee  We can then use
(\ref{invzwbn}) to compute $1/j$ corrections to the inverse zweibein (\ref{loizwbn})
\beqa h_1
&=&\;\;\frac 14
{\sqrt{1+|\zeta|^2}}\;\;\biggl\{2+|\zeta|^2+\frac1{16j}(8+32|\zeta|^2
+13|\zeta|^4)\cr& &\cr& &\qquad\qquad\qquad\quad
-\;\;\bar\zeta^2\biggl(1+\frac1{16j}(6+ 17|\zeta|^2)\biggr)\biggr\}+{\cal
  O}(1/{j^{2}})
\cr
& &\cr
h_2&=&-\frac i4 {\sqrt{1+|\zeta|^2}}\;\;\biggl\{2+|\zeta|^2+\frac1{16j}(8+32|\zeta|^2
+13|\zeta|^4)\cr& &\cr& &\qquad\qquad\qquad\quad
+\;\;\bar\zeta^2\biggl(1+\frac1{16j}(6+ 17|\zeta|^2)\biggr)\biggr\}+{\cal
  O}(1/{j^{2}})
\;\eeqa
For the coefficients (\ref{cfofcb}) and (\ref{cfofqt}) of the cubic
and quartic terms, respectively, of scalar field
theory we get
\beqa  G^{\zeta,\zeta\zeta}&=&\frac1{8j}\;\zeta^3\;(1 + |\zeta|^2)^2\;(3 +
        2|\zeta|^2)\cr
& &\cr
 G^{\bar\zeta,\zeta\zeta}&=&\frac1{8j}\;\zeta\;(1 + |\zeta|^2)^2\;(4 + 6|\zeta|^2+ 3|\zeta|^4)\cr & &\cr
 G^{\zeta,\zeta\bar\zeta}&=&\frac1{8j}\;\zeta\;(1 + |\zeta|^2)^3\;(4 +
        3|\zeta|^2) \cr & &\cr G^{\zeta\zeta,\bar\zeta\bar\zeta}&=&
\frac1{8j}\;(1 + |\zeta|^2)^3\;(2 +
        2|\zeta|^2 + |\zeta|^4) \cr
& &\cr
 G^{\bar\zeta\bar\zeta,\zeta\bar\zeta}&=&
\frac1{8j}\;\bar\zeta^2\;(1 + |\zeta|^2)^3\;(2 +|\zeta|^2)\;,
 \eeqa along with their complex conjugates.  These results looks quite
 different from the leading cubic and quartic terms obtained in
 (\ref{neqlone}) from the Voros star product.

\section{Concluding Remarks}
\setcounter{equation}{0}

Here we remark on possible generalizations of this work.

Although messy it is straightforward to go beyond the first order
corrections.  Three different contributions occur at second order in
$\theta_0$.  They are: ${\cal S}^0_2$, ${\cal S}^1_1$, ${\cal S}_0^2$.
The first term is star product independent, while the last requires
expanding the star product to second order.  Also one can try other
star products on the plane, such as those developed in \cite{Hammou:2001cc},
\cite{Lubo:2004mz}.  As noted previously, our procedure  requires  a star product written
directly on  the coordinate patch.

 Another obvious generalization is to go to more than two dimensions.
Since we need a nonsingular Poisson structure we should restrict to an
even number of dimensions ${\tt d}$.  For ${\tt d}>2$  we
must distinguish between the set of all orthogonal frames $\{ {\cal
O}\}$ and the set of orthogonal frames, which we denote   by $\{\tilde
{\cal O}\}$, with a constant Poisson structure, i.e. the Poisson
tensor has constant components $\tilde\theta^{ab}$. We cannot identify
$\{ {\cal O}\}$ with  $\{\tilde {\cal O}\}$ because the constant
Poisson tensor
 is  not  preserved under general local orthogonal transformations
(\ref{lotrnsfmtn}).   Instead   $\{\tilde {\cal O}\}\subset\{ {\cal
  O}\}$, and the two sets of frames are related by local othogonal
 transformations. Now denote by $\tilde e_{\;\;\mu}^c(x)$ those
 veilbeins which  transform from the coordinate patch  ${\cal P}$ to $\tilde {\cal
 O}$.  For dimension greater than two, the set of such veilbeins
 $\{\tilde e_{\;\;\mu}^a(x)\}$ is a subset of  the set of all
 veilbeins $\{ e_{\;\;\mu}^a(x)\}$. If  the basis vectors of
 $\tilde{\cal O}$ are $\frac{\partial}{\partial \tilde y^a}$, then
\be \frac{\partial}{\partial x^\mu} = \tilde e_{\;\;\mu}^a(x)\;
\frac{\partial}{\partial\tilde y^a}\label{rltnbtwnortbsaf}\ee Since we again need
coordinate maps to flat noncommutative manifolds, the frames $\{
\tilde{\cal O}\}$  need to be
 coordinate bases.  Calling its coordinates $\tilde y^a$, \be
\{ \tilde y^a,\tilde y^b\}=\tilde\theta^{ab} ={\rm constants}\ee As
$\tilde\theta^{ab}$ is nonsingular, using (\ref{rltnbtwnortbsaf}), we
get
\be \frac{\partial}{\partial x^\mu} =\tilde e_{\;\;\mu}^a(x)\;[\tilde\theta^{-1}]_{ab}\;\{\tilde y^b\;,\;\;\}\label{frddx}
\ee  Then the  action of a massless scalar field $\phi$
in a local region $\sigma$ of ${\mathbb{R}}^{\tt d}$ is
\be {\cal S}_0=\frac 1{2}\int_\sigma d^{\tt d}y \;[\tilde\theta^{-1}]_{ab}\;
\;[\tilde\theta^{-1}]_{ac}\;\{\tilde y^b\;,\phi\}\;
\{\tilde y^c\;,\phi\}
\;,\ee
which is easily generalized to the noncommutative case.  It remains to
find the maps from ${\bf x}^\mu$ to $\tilde {\bf y}^a$, the
noncommutative analogues of $x^\mu$ to $\tilde y^a$, respectively, and
re-express the noncommutative action in terms of the symbols of ${\bf
x}^\mu$.

It would also be of interest to go beyond scalar field
theories. In addition to including a mass and interaction term in the
scalar field theory  is the possibility of changing the target space.
For example, one can investigate fuzzy corrections to the nonlinear
$\sigma$-model and its soliton solutions. A more challenging
generalization involves the inclusion of spin. For this we need the
analogue of a spin connection $[\omega_\mu]^a_{\;\;b}(x)$. The
covariant derivative of components $u
^a$ of a vector $V$ in a local orthogonal frame  ${\cal O}$ is given by
\be
D_\mu u^a = \frac{\partial}{\partial x^\mu} u^a
+[\omega_\mu]^a_{\;\;b} u^b\; \ee Upon transforming to another local
orthogonal frame ${\cal O}'$
 \be D_\mu u^a\rightarrow[D_\mu u]'^a = \lambda^a_{\;\;b}(x)
D_\mu u^b\;,\ee where $[\omega_\mu]^a_{\;\;b}(x)$ transforms as \be
\omega_\mu\rightarrow \omega_\mu' = \lambda \omega_\mu \lambda^{-1} -
\frac{\partial}{\partial x^\mu}\lambda\; \lambda^{-1} \ee
So here in addition to the  noncommutative maps, which play the role
of noncommutative veilbeins, one needs the noncommutative analogue of
the spin connections.  The latter requirement should be similar to
having a Dirac operator.

Up to now we have  examined a single
coordinate patch.  It is natural to ask whether the full noncommutative
manifold can be described in terms of coordinate patches.  For this it
is  necessary to define the analogue of transition functions on
overlapping
patches.  It is possible that the procedure can be used to define new
noncommutative manifolds.

A final possibility we mention is to make the analogue of the
veilbeins and spin connection dynamical and thus move in the direction
of a noncommutative general relativity.  A step in this direction is
to consider dynamical  maps from operators ${\bf x} ^i $ satisfying an
arbitrary noncommutative algebra to flat noncommutative manifolds
spanned by operators ${\bf y} ^i $.  This may be facilitated as in
\cite{Jackiw:2004nm} with the introduction of dynamical fields
$A_i({\bf x})$, writing the map as ${\bf x}^i= {\bf y} ^i
+\theta^{ij}A_j({\bf x})$.   The goal would then be to integrate over
all maps, i.e. fields $A_i({\bf x})$, and spin connections
$[\omega_\mu]^a_{\;\;b}(x)$ or Dirac operators.

\bigskip
\bigskip

${\bf Acknowledgement}$

  We are very grateful to F. Lizzi, D. O'Connor  and P. Vitale for useful
discussions.  A.S. wishes to thank G. Marmo and members of the theory
group of the University of Naples for their warm hospitality while
this work was being completed.  This work was supported in part by the
joint NSF-CONACyT grant E120.0462/2000 and  DOE grants
DE-FG02-85ER40231 and DE-FG02-96ER40967.

\end{document}